\documentclass[journal]{IEEEtran}
\usepackage{amsmath,amsfonts}
\usepackage{algorithmic}
\usepackage{algorithm}
\usepackage{array}
\usepackage[caption=false,font=normalsize,labelfont=sf,textfont=sf]{subfig}
\usepackage{textcomp}
\usepackage{stfloats}
\usepackage{url}
\usepackage{verbatim}
\usepackage{graphicx}
\usepackage{cite}
\usepackage{orcidlink}
\usepackage{booktabs}
\usepackage[table]{xcolor}
\usepackage{xcolor}
\usepackage{balance}
\usepackage{tabularx}
\usepackage{ragged2e}

\begin{document}

\title{From Agentic to Autogenic Network Management for AI-Native 6G and Beyond: A Standards Perspective}


\author{Petar Djukic,~\and~Sudipta Acharya,~\and~Takai Eddine Kennouche,\and~and Burak Kantarci
\thanks{P. Djukic is with Bell Labs Research, 600 March Rd, Kanata, ON K2K 2E6. (e-mail: petar.djukic@nokia-bell-labs.com).}
\thanks{T. Kennouche is with Nokia Technology Standards, Nokia France, 12 rue Jean Bart Massy 91300. (e-mail: takai.kennouche@nokia.com)}
\thanks{S. Acharya and B. Kantarci are with University of Ottawa, 350 Legget Drive, Kanata, ON, Canada, K2K 2W7 (e-mail:\{sacharya2,burak.kantarci\}@uottawa.ca)}
}



\maketitle

\begin{abstract}
Standards bodies, including TM Forum, 3GPP, and ETSI, are converging on Agentic AI as the foundation for next-generation network management, where Large AI Model (LAM)-based agents autonomously interpret intent, coordinate resources, and adapt operational behaviors at runtime. However, achieving this vision at the scale and complexity of 6G networks requires management systems that can generate and evolve their own automation software during operation. We introduce \textit{Autogenic network management}, a reference architecture that extends agentic capabilities with self-programming, self-reflection, self-orienting, and self-architecting capabilities. The architecture supports practical staged deployment beginning with human-supervised LAM-based agents and progressing toward autonomous operation as confidence builds. We demonstrate the approach through high-priority operator scenarios drawn from TM Forum's autonomous network use cases, showing how autogenic management addresses real operational challenges. We conclude with a research roadmap outlining the technical advances needed to make autogenic network management realistic in future 6G networks.

\end{abstract}
\pagestyle{empty}
\thispagestyle{empty}

\section{Introduction}\label{section:introduction}

\IEEEPARstart{T}{he} next generation of 6G promises to be AI-native, embedding machine learning (ML) across every layer, function, and interface \cite{3gpp-tr-22870}. While this vision offers unprecedented capabilities, from semantic communications to intent-driven operations, it also introduces operational complexity that exceeds current network management approaches. As ML software proliferates throughout network functions (NFs), maintaining performance at scale while managing interdependencies becomes a critical challenge.

This paper argues that achieving autonomous AI-native operations at scale requires autonomous network management where the management plane can generate new automation software, validate its correctness, and modify its own operational structure during runtime. We term this form of autonomy \emph{``autogenic"} and present \emph{autogenic network management} as an architectural approach for realizing Agentic AI in 6G networks, where LAM-based agents autonomously make decisions, generate solutions, and evolve system behavior in response to operational requirements. 

In this article, we present the architectural foundations, design principles, and a research roadmap for autogenic network management systems. Our work directly addresses the architectural challenges identified by standards bodies for agent-based network management, providing a reference architecture compatible with TM Forum's autonomous network framework and ETSI's experiential networked intelligence vision. 

\subsection{How Do AI-Native Networks ``Think''?}\label{think}

In a future 6G system, artificial intelligence is expected to be pervasive. ML models, adaptive controllers, and data-driven optimization routines will operate across all layers and interfaces. The air interface will incorporate AI-native mechanisms for detection, scheduling, and other core Radio Access Network (RAN) operations. LAMs will coordinate functions across the network \cite{3gpp-tr-22870, etsi-gr-eni-051}.

This level of AI integration introduces a new class of operational challenges. For example, electricity over-consumption in the network can impact sustainability targets and increase operational expenditure. Addressing such conditions through manual diagnostics becomes impractical at scale. Instead, operators will require management systems that accept high-level intents, such as sustainability objectives, and translate them into appropriate corrective strategies without explicit procedural instructions.

In the electricity over-consumption scenario, the management system responds to increased electricity use by formulating an investigation objective and analyzing electricity consumption trends across the RAN. Correlation with key performance indicators reveals that the ML model used for Medium Access Control (MAC) scheduling has experienced drift. As a result, the scheduler selects suboptimal transmission decisions that increase retransmissions and keep radios active for longer intervals. This condition was not captured during design or initial validation; the network identifies the underlying cause through its own runtime analysis.

\IEEEpubidadjcol

Having identified model drift as the cause, the network must address it. Retraining the MAC scheduler is dynamic and may require the system to self-architect its training pipeline, adopt new data sources, or design novel feature-extraction routines. It must self-program data collection, federate across domains, and self-program the training procedure by orchestrating resources and tuning algorithms. After training a new model, the network must self-reflect by validating the model, running safety checks, and monitoring for unintended consequences.

In AI-native 6G networks, this \emph{detect $\rightarrow$ diagnose $\rightarrow$ retrain $\rightarrow$ validate $\rightarrow$ deploy $\rightarrow$ monitor} cycle should continuously repeat across many network, operational, and service domains \cite{tmf-ig1251}. Manual intervention and human oversight do not scale, as the complexity and pace of change exceed human operational capacity. A trusted autonomous management system must manage this lifecycle at scale and will be essential for 6G success.
\begin{figure*}
\centering
\includegraphics[width=6.3in]{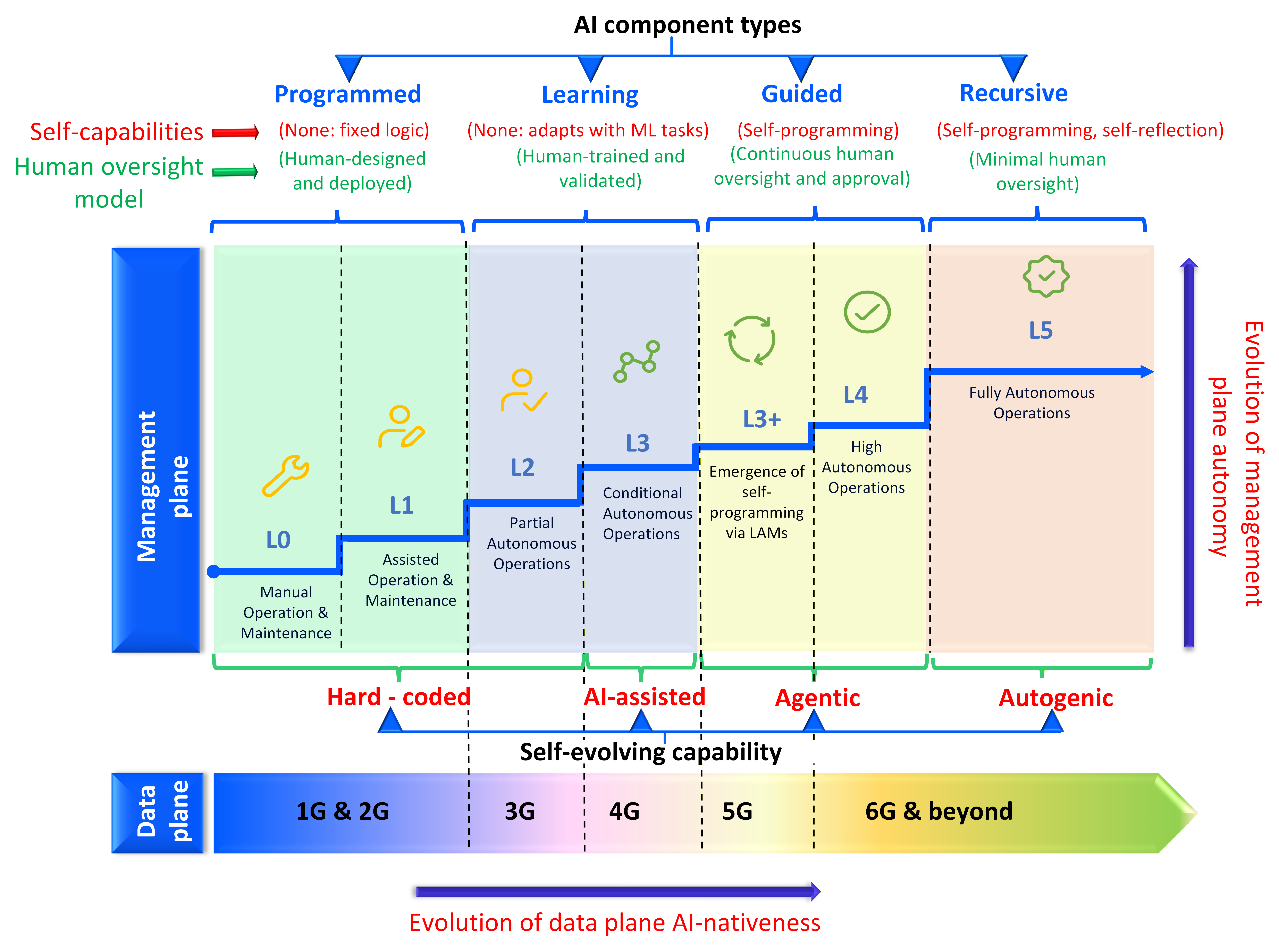}
\caption{Relationship between AI component types, management-plane autonomy levels, and their co-evolution with data-plane AI-nativeness. Programmed → learning → guided → recursive components correspond to L0–L5 autonomy in the management plane. The figure also highlights the increasing degree of self-evolving capability in the management plane (hard-coded → AI-assisted → agentic → autogenic) and the parallel evolution of the data plane from non-AI to AI-native behavior. \textbf{Color coding:} Management plane → Different AI component types; Data plane → Different network generations.}
\label{figures:evolution}
\end{figure*}
\subsection{Strategic Business Drivers}

6G enables revenue streams that are not feasible today. For example, immersive extended reality requires ultra-low latency and high reliability, while industrial automation requires precise timing and deterministic communications for safety-sensitive operations. These applications create new market segments and premium service tiers. At the same time, AI-native networks should lower infrastructure costs because data-driven ML software runs on commodity hardware. Additional savings in software infrastructure come from automated generation of data-driven ML-based algorithms during network operations, eliminating the cost of manual algorithm design.

In AI-native networks, data-driven automation must be created within the network itself, as that is where the operational data resides. As networks continue to scale and diversify, the cost of human oversight, troubleshooting, and configuration becomes unsustainable, making autonomous operation essential for managing operational expenses. Network management standards reflect these business drivers. Standards bodies are now considering automation architectures in which LAM-driven automation underpins next-generation network management \cite{etsi-gr-eni-051} with a shared architectural language and LAM-based agents as its foundational components\cite{tmf-ig1251, tmf-ig1251c, tmf-ig1251d}.

\subsection{The Hidden Challenges}

The AI-native network vision promises networks that are more efficient, responsive, and capable of supporting new services. As the industry moves from theory to deployment, practical challenges emerge. The complexity and scale of 6G requirements reveal limitations that are both technical and systemic.

Hidden problems stem from the proliferation of ML software. Unlike traditional software, ML software erodes modular boundaries and creates entangled dependencies through data, features, and feedback loops. Risks such as the Changing Anything Changes Everything (CACE) principle, where seemingly minor changes to one ML model can unexpectedly degrade performance across the entire system due to hidden interdependencies, accumulate quietly and often surface only after deployment and at scale \cite{sculley2015hidden}.

As more of the responsibility for maintaining the performance of ML models moves into network operations, the Mobile Network Operators (MNOs) must change their operational practices to include more in-network software creation, validation, and testing. This is where the need for a new type of management plane becomes apparent, prompting us to propose the \emph{Autogenic network management} system.

\subsection{Key Concepts}

The key to scalable AI-native autonomous networks lies in the capability emerging from LAM research. We define the architectural concepts here and explain how they build upon each other.

\begin{itemize}
    \item \emph{Autonomous networks} are networks that can perceive their environment, make decisions, and take actions to achieve predefined operational goals with minimal human intervention. They operate recursively through many autonomous systems spanning network, operational, and service domains, individually progressing through maturity levels (L0-L5) based on their degree of independence from human oversight \cite{tmf-ig1251}.

\item \emph{Agents} are software entities that act autonomously or semi-autonomously to perform tasks, make decisions, and interact with other agents or systems \cite{etsi-gr-eni-051}. In network management, LAM-based agents can interpret intent, generate procedural knowledge at runtime, and coordinate resources to drive autonomous adaptation \cite{tmf-ig1251d}. The guided and recursive components introduced in this paper represent two implementation approaches to Agentic AI, differing in their degree of autonomous decision-making and validation.

\item \emph{Autogenic systems} represent a novel architectural pattern. These systems can generate their own goals, executable behaviors, and structural control logic through four capabilities: \emph{self-orienting} (generating new objectives), \emph{self-programming} (synthesizing new behaviors), \emph{self-reflection} (evaluating reasoning processes), and \emph{self-architecting} (transforming system structure). We introduce autogenic systems as a general concept applicable to any software-centric domain but demonstrate their specific application to network management.

\item \emph{LAM-enabled Self-evolving networks} apply the autogenic systems pattern to network management. They are networks whose management plane implements autogenic capabilities through LAMs, enabling software components and behaviors to evolve during runtime in response to changing requirements, traffic patterns, and environmental conditions. This differs from the pre-LAM self-evolving network concept \cite{darwish2021vision}, which focused on ML-driven network optimization and resource management without the capability to generate new goals or modify the system architecture during runtime.
\end{itemize}

\section{AI Component Spectrum and LAM-Enabled Autonomy}\label{section:lam-enabled-network-management}




To realize the full benefits of automation, networks must operate at L3+ autonomy levels \cite{tmf-ig1251} (Fig. \ref{figures:evolution}), where network management handles most operational situations. Achieving this through manually developed, design-time software is costly and difficult to scale.

\begin{figure}[!t]
\centering
\includegraphics[width=3.5in]{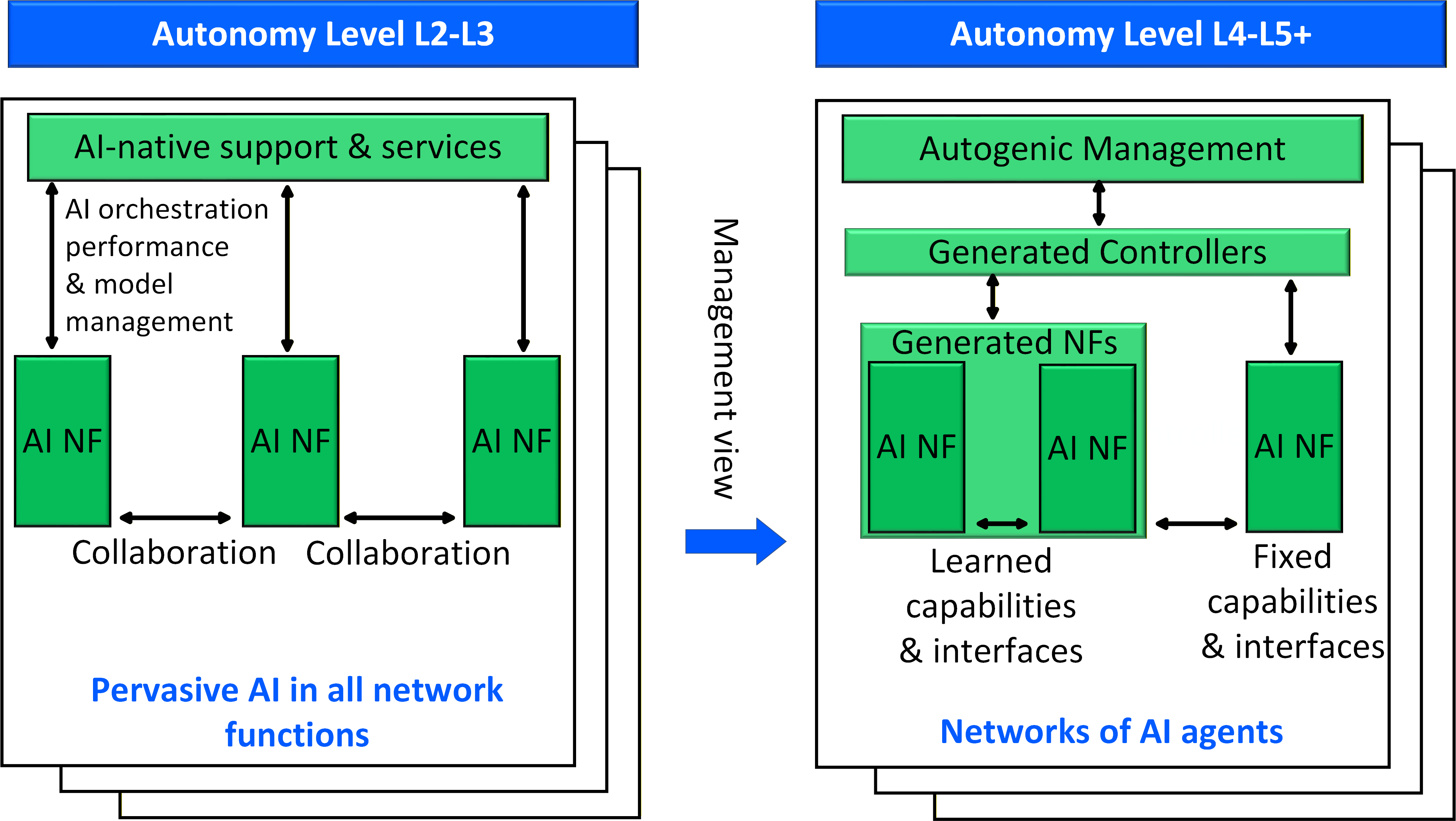}
\caption{Autonomy-level view of AI-native networks. At autonomy levels L2--L3 (left), AI is pervasive across network functions and orchestration, enabling performance monitoring and model management while capabilities and interfaces remain fixed. At autonomy levels L4--L5+ (right), autogenic management generates controllers that create NFs with learned capabilities and interfaces, enabling structural self-evolution while coexisting with fixed AI/NFs.}
\label{figures:self-evolving-networks}
\end{figure}

Although AI-native networks improve resource optimization and event prediction, ML components alone do not create, maintain, or evolve the management software required for network operation. However, managing model lifecycles, performance drift, and cross-domain dependencies demands continuous software evolution. Autogenic network management addresses this challenge by shifting software creation into network operations, enabling the management plane to generate and evolve automation at runtime (Fig. \ref{figures:self-evolving-networks}).

LAMs enable this approach through self-programming, generating procedural logic and predicting its execution \cite{anthropic2024claude}. While LAMs are \emph{partial} autogenic systems, an autogenic architecture provides the autonomy required to scale management of AI-native networks. We therefore advocate integrating LAMs within an architecture that makes the overall system autogenic. Combining LAMs with autogenic capabilities yields \emph{LAM-enabled self-evolving networks}, in which management-plane software components and behaviors evolve at runtime.

This view of the management plane is reflected in recent standards efforts \cite{3gpp-tr-22870,etsi-gr-eni-051, tmf-ig1251, tmf-ig1251c, tmf-ig1251d}, which recognize both the strengths of ML for domain-specific optimization and the challenges of increasing system complexity. 3GPP defines AI-native 6G requirements \cite{3gpp-tr-22870}, ETSI investigates AI agents for next-generation network slicing \cite{etsi-gr-eni-051}, and TM Forum advances agent-based architectures for L4+ autonomous networks \cite{tmf-ig1251, tmf-ig1251c, tmf-ig1251d}. Collectively, these efforts point toward LAM-enabled network management that extends beyond predefined optimization to runtime self-evolution.

\subsection{The Spectrum of AI Components in AI-Native Networks}\label{componants}

AI-native networks will be built with a spectrum of component types, each with different capabilities for processing, adaptation, and evolution, as illustrated in Fig. \ref{figures:evolution}. The figure shows how AI component types relate to management-plane autonomy and data-plane AI-nativeness. 

\emph{Programmed}, \emph{learning}, \emph{guided}, and \emph{recursive} components align with autonomy levels L0-L5 in the management plane. They also correspond to increasing degrees of self-evolving capabilities: hard-coded → AI-assisted → agentic → autogenic. At the same time, the data plane evolves from non-AI to AI-native behavior across network generations (1G–6G and beyond). The figure highlights that autonomy growth in the management plane and AI-nativeness growth in the data plane progress together, with guided and recursive components enabling agentic and autogenic behavior, respectively. The AI component types are detailed below.

\emph{Programmed components} are the primary means by which networks are built today. They are designed, implemented, and deployed entirely by humans, who define all operational logic in advance. They execute procedural logic specified at design time, providing deterministic but non-adaptive behavior. These components enable the most basic, i.e., L0-L1 network autonomy.

\emph{Learning components} implement ML-based predictions. Humans train and validate the models, while the models execute autonomously within their trained tasks, such as traffic prediction, anomaly detection, or resource optimization. In AI-native 6G networks, these components enable L2--L3 autonomy through automated pattern detection for specific ML tasks \cite{tmf-ig1251}. However, they cannot interpret intent or generate new behaviors, which are required for L4 autonomy.

Based on recent advances in LAMs, we also recognize two emerging types of components. Unlike programmed or learning components, which have fixed logic or adapt within predefined ML tasks, these components have strong self-programming capabilities.

\emph{Guided components} are LAM-based semi-autonomous agents with self-programming capabilities. They require \emph{human-in-the-loop oversight} where operators supervise workflows, validate generated code, and intervene in exceptions. Guided components interpret intent and generate new code, enabling operators to create programmed components during network operations. However, humans remain part of the automation creation process, reviewing and approving generated components. The runtime code generation capability provided by guided components is key to enabling the TM Forum L4 autonomy vision at scale  \cite{tmf-ig1251}.

\emph{Recursive components} are advanced LAM-based agents with self-programming and self-reflection capabilities. Self-reflection enables them to evaluate their own outputs, detect errors, and improve their code autonomously. Unlike guided components, recursive components can create new components and validate them without human approval. This eliminates the human bottleneck in scaling network management automation.

Autogenic architectures use recursive components to enable self-orienting and self-architecting capabilities, extending autonomy beyond L4 (Section~\ref{section:autogenic-network-management}). Only guided and recursive components qualify as agents because self-programming allows them to interpret intent and generate behaviors at runtime. Current systems primarily rely on guided components with human oversight, but are expected to evolve toward recursive components capable of autonomously creating and validating automation with minimal supervision.


\subsection{Evolution Between Levels of Network Automation}

Fig.~\ref{figures:self-evolving-networks} contrasts pervasive AI deployment at L2–L3 with networks of AI agents at L4–L5+. The key architectural shift is from orchestrating fixed components to generating new components at runtime. LAM-enabled self-evolving networks extend automation by enabling software agents to adapt, generate, and evolve management components with minimal human intervention. In the medium term, operators will rely on guided components to reach L4 autonomy, while in the longer term, recursive components will enable L5 and L5+ autonomy.

The LAM-enabled self-evolving networks perspective highlights a shift in standardization from defining functionality to enabling evolution. Instead of specifying fixed behaviors, standards would define how agents, both guided and recursive, generate and validate new behaviors during operation. Progressing toward L5 and beyond requires standardized autogenic architectures in which recursive components can formulate goals, synthesize implementations, and self-validate with reduced human supervision.

\section{Autogenic Network Management}\label{section:autogenic-network-management}

As 6G networks progress toward a higher level of autonomy, the ability to adapt, learn, and improve during operation becomes necessary. This section outlines the architectural shift to autogenic network management, required for networks to operate intelligently and to evolve over time. Autogenic network management leverages guided and recursive components (as discussed in Section \ref{componants}) to achieve self-orienting and self-architecting capabilities. The reference architecture of the proposed autogenic network management system, as shown in Fig. \ref{figures:autogenic-system-reference-architecture}, illustrates how organizing recursive components within specific subsystems creates these system-level capabilities.

\subsection{Why ``Autogenic''?}

We use the term “autogenic” to emphasize the generative capability that distinguishes autogenic network management from the current view of autonomous networks. The term refers to systems that can generate their own actions, introduce adaptations, and evolve.

\subsection{Autogenic Capabilities}

Autogenic systems comprise four core capabilities that enable self-evolution, as follows.
\begin{enumerate}
    \item \emph{Self-programming}: The system synthesizes new behaviors autonomously. Rather than only tuning parameters or reconfiguring programmed and learning components, self-programming allows the system to generate procedural logic and realize novel objectives or redefine objectives in response to unexpected conditions.
    \item \emph{Self-reflection}: The system evaluates its own reasoning processes and decision frameworks. Beyond measuring performance against fixed criteria, self-reflection also examines and adapts the criteria, enabling more robust adaptation.
    \item \emph{Self-orienting}: The system generates new objectives rather than only selecting from predefined goals. Where conventional systems optimize and adapt within fixed objectives and constraints, self-orienting systems redefine objectives and create new objectives when required.
    \item \emph{Self-architecting}: The system modifies its core architecture and integrates new functionality at runtime. This includes synthesizing interface adaptors, modifying protocols, and restructuring internal components to meet novel requirements.
\end{enumerate}
\begin{figure}[!t]
\centering
\includegraphics[width=3.5in]{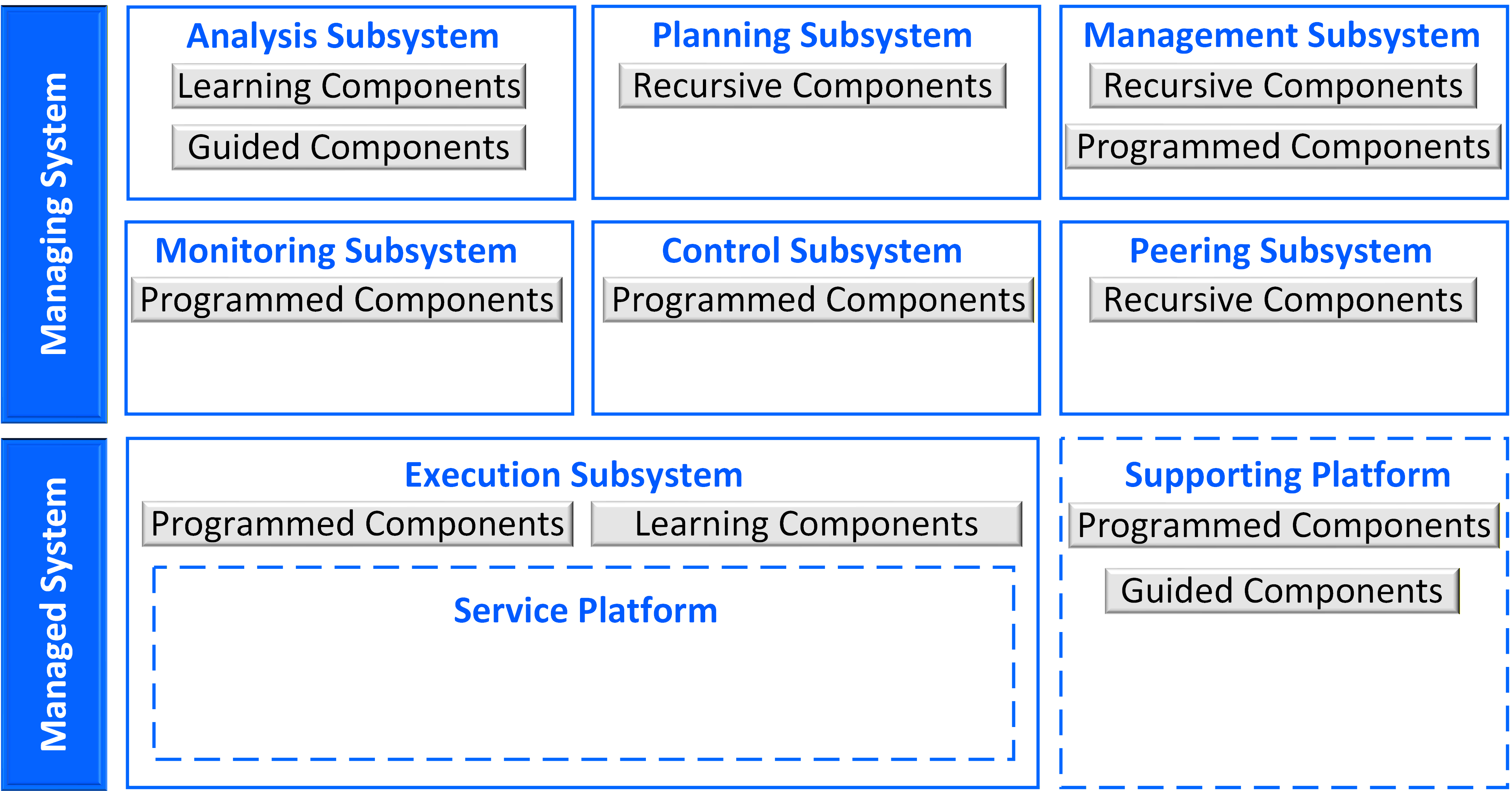}
\caption{Autogenic systems are organized into seven subsystems on a Supporting Platform. Execution delivers services to external users. Monitoring, Analysis, Planning, and Control implement AADE loops over the Execution subsystem. Management governs the evolution of the system. Peering coordinates with external systems.}
\label{figures:autogenic-system-reference-architecture}
\end{figure}
\subsection{Reference Architecture of the Autogenic System}

In Section~\ref{componants}, we have introduced four component types in AI-native networks. In this section, we establish different subsystems within autogenic system and show how they interact.

The architecture exposes a dual scope of operation: external and internal. The external scope of operations is handled by a managed system, while the internal scope of operations is handled by the managing system. The managed system is the Execution subsystem, which delivers services to external users. The other six subsystems comprise the managing system: Monitoring collects telemetry, Analysis detects patterns and anomalies, Planning generates adaptation strategies and synthesizes new behaviors, Control enforces decisions and applies changes. The Peering subsystem enables federated autonomy through coordination with external systems, while the Management subsystem governs system evolution, setting long-term goals and validating structural and behavioral changes.

The Planning and Control subsystems extend traditional management beyond AADE (Awareness, Analysis, Decision-making, Execution) loops \cite{tmf-ig1251} by governing the managing system itself. They generate, evolve, and retire components while maintaining structural cohesion. The Supporting platform represents the hardware and software on which the AI-native network and its management run.

Having a reference architecture provides several benefits. First, it shows how component-level capabilities (self-programming, self-reflection) can compose into system-level capabilities (self-orienting, self-architecting). Second, functional separation enables reasoning about safety, security, trust, and verification within specific subsystems rather than across the entire system. Third, the architecture groups components by their time scales, allowing subsystems to operate independently and synchronize through well-defined coordination mechanisms.

The Execution subsystem operates at service delivery speeds, requiring deterministic behavior from programmed and learning components. The Monitoring, Analysis, Planning, and Control subsystems operate at adaptation time scales, implementing AADE loops that reason about system state. The Management subsystem operates at evolution time scales, where new components are generated to change the system.

\textcolor{black}{This organization maps AI capabilities to architectural responsibilities. Self-programming is realized in Planning and Management through the synthesis of adaptation strategies, behaviors, and components. Self-reflection is supported by Analysis and Control through state evaluation, action validation, and outcome monitoring. Self-orienting and self-architecting are primarily governed by Management, which oversees objectives, component relationships, and structural evolution.}

\subsection{Enabling Self-Orienting and Self-Architecting Capabilities}

The reference architecture enables self-orienting and self-architecting by deploying recursive components within the managing system.

Self-orienting emerges when recursive components in Planning and Management subsystems use self-programming to synthesize new objective functions based on observed network behavior. Rather than selecting from predefined goals, these components generate novel operational objectives when environmental conditions or requirements change.

Self-architecting occurs through the Management subsystem, where recursive components modify the system's structural organization. Using self-programming, these components can synthesize interface adaptors, restructure component relationships, and integrate new functionality at runtime. The dual-scope principle is critical here: the Management subsystem applies these capabilities not only to the managed system (the AI-native network) but also to the managing systems themselves, enabling open-ended architectural evolution.

These capabilities represent an active research frontier. While today's prototypes demonstrate goal generation and structural modification in constrained domains, achieving reliable self-orienting and self-architecting in production networks requires advances in all four autogenic capabilities and safety frameworks (as discussed in Section~\ref{section:challenges-and-roadmap}).

Current LAMs approximate recursive behavior through iterative code generation and testing \cite{anthropic2024claude}, but lack formal correctness guarantees. Guided components address this limitation through human-in-the-loop validation, providing safety while retaining self-programming capabilities. The proposed architecture supports a gradual transition from guided to recursive components as verification frameworks mature.
\begin{table*}[!t]
\centering
\caption{High-Value Scenarios for Autonomous Networks}
\label{table:high-value-scenarios-for-autonomous-networks}

\renewcommand{\arraystretch}{1.25}
\setlength{\tabcolsep}{7pt}

\rowcolors{2}{green!8}{blue!5}

\begin{tabularx}{\textwidth}{
>{\centering\arraybackslash}p{0.05\textwidth}
>{\RaggedRight\arraybackslash}p{0.22\textwidth}
>{\RaggedRight\arraybackslash}X
}
\toprule
\rowcolor{blue!18}
\textbf{No.} & \textbf{Scenario} & \textbf{Description} \\
\midrule

\textbf{1} & \textbf{Service Marketing} 
& Autonomous discovery and qualification of service opportunities; dynamic, intent-driven campaigns. \\

\textbf{2} & \textbf{Service Provisioning} 
& Automated fulfillment and activation of services, including resource allocation and configuration. \\

\textbf{3} & \textbf{Service Assurance} 
& Continuous monitoring and assurance of service performance, quality, and reliability. \\

\textbf{4} & \textbf{Complaint Handling} 
& Autonomous management of customer complaints, root cause analysis, and proactive resolution. \\

\textbf{5} & \textbf{Network Planning} 
& Cognitive, data-driven planning and design; capacity forecasting and scenario analysis. \\

\textbf{6} & \textbf{Network Deployment} 
& Automated rollout, configuration, and integration of new network functions and infrastructure. \\

\textbf{7} & \textbf{Fault Management} 
& AI-enabled, closed-loop fault detection, diagnosis, and autonomous rectification. \\

\textbf{8} & \textbf{Network Change} 
& Zero-touch change management; impact analysis, validation, and zero-outage updates. \\

\textbf{9} & \textbf{Quality Optimization} 
& Continuous, closed-loop optimization of network quality; real-time performance tuning. \\

\textbf{10} & \textbf{Energy Efficiency Optimization} 
& AI-driven optimization of energy consumption; predictive power saving and dynamic adjustment. \\

\textbf{11} & \textbf{Resource Management} 
& Autonomous management and optimization of network resources; dynamic allocation and scaling. \\

\bottomrule
\end{tabularx}
\end{table*}

\begin{figure}[!htb]
\centering
\includegraphics[width=3.5in]{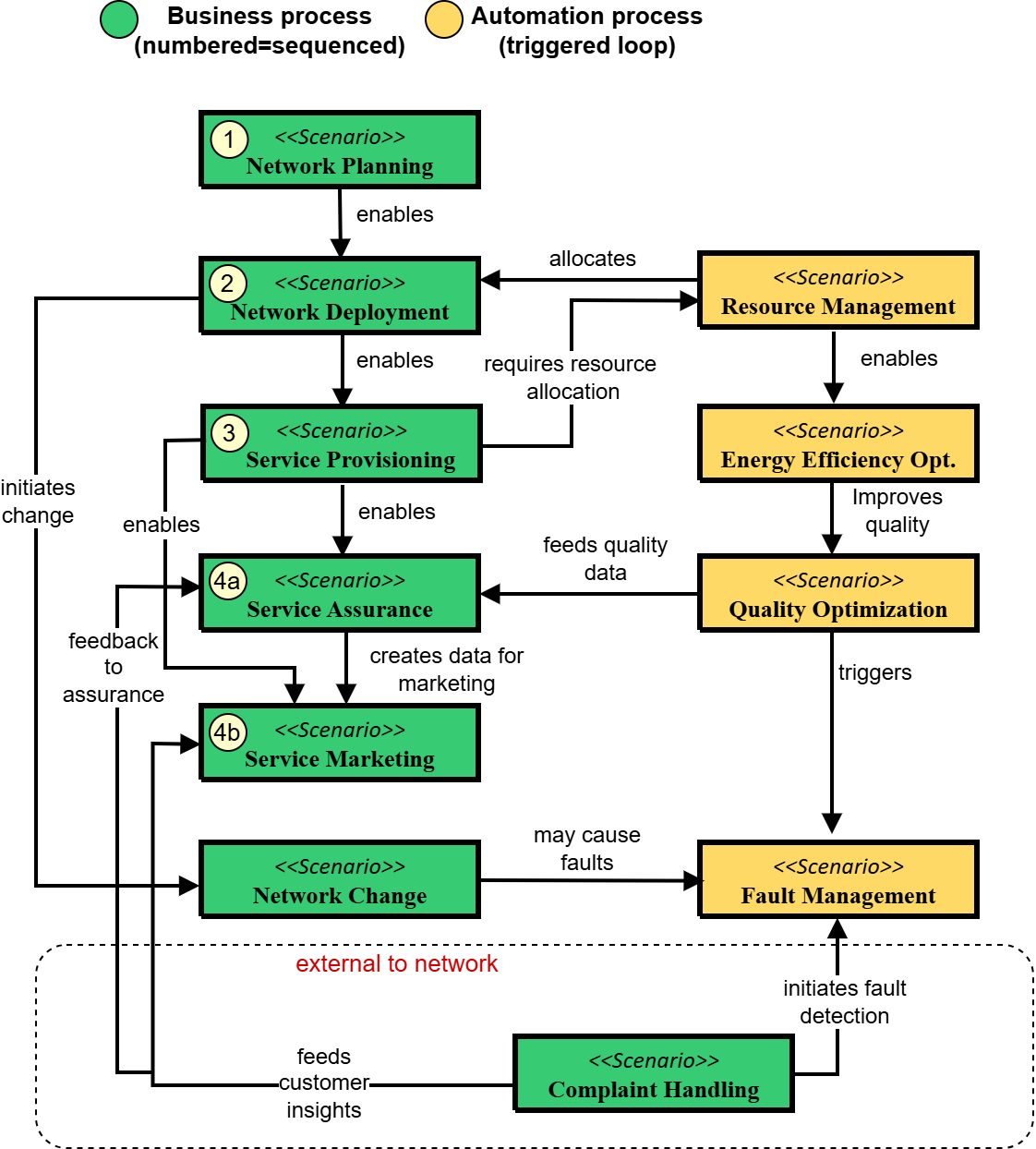}
\caption{\textcolor{black}{High-value scenario (as listed in Table \ref{table:high-value-scenarios-for-autonomous-networks}) relationships. Business processes (green) follow a numbered precedence order: Network Planning (1) enables Network Deployment (2), which supports Service Provisioning (3), concurrently enabling Service Assurance (4a) and Service Marketing (4b). Automation processes (amber yellow) are triggered control loops attached as feeder/trigger inputs and left unnumbered; for example, Service Assurance feeds data into Quality Optimization, which can trigger Fault Management when it detects issues. Complaint Handling, triggered externally by the customer, is shown below the network boundary. This interconnectedness reflects that autonomous networks must coordinate multiple workflows simultaneously to maintain service continuity and improve performance.}}
\label{figures:high-value-scenarios-usecase-diagram}
\end{figure}

\subsection{Design Principles}

Six principles guide autogenic system design:
\begin{enumerate}
\item \emph{Minimal Viable Complexity}: Use the simplest component type needed for each subsystem.
\item \emph{Dual-Scope Management}: Separate service delivery from adaptation and evolution.
\item \emph{Reuse Over Reinvention}: Build on proven operational components.
\item \emph{Self-Contained Resilience}: Implement appropriate resilience mechanisms for each complexity level.
\item \emph{Fallback Readiness}: Enable safe degradation when recursive components are unavailable.
\item \emph{Component-Only Construction}: Build all subsystems from modular components.
\end{enumerate}

\subsection{Autogenic Management of 6G AI-Native Networks}

In 6G networks, the RAN and Core Network (CN) are the managed systems. The subsystems drive adaptation, optimization, and resilience across RAN and CN components. The management system implements AADE loops for the Execution subsystem, improving network efficiency and reducing infrastructure costs through continuous optimization. The Management system applies autogenic capabilities to itself, improving automation software creation and reducing operational costs by minimizing human intervention.

\section{Network Operator Perspective}\label{section:network-operator-perspective}

This section demonstrates autogenic network management through operational scenarios that 6G networks must support. We examine one of these scenarios- \emph{fault management}, in detail, to show how agents coordinate to resolve network issues autonomously.

\subsection{Operational Scenarios}

Table~\ref{table:high-value-scenarios-for-autonomous-networks} lists 11 high-value scenarios identified by TM Forum that define the operational scope of autonomous networks \cite{tmf-ig1251c}. Fig.~\ref{figures:high-value-scenarios-usecase-diagram} shows the relationships between the scenarios. While the scenarios were defined in the context of L4 autonomy with guided components, they still remain valid for higher levels of autonomy envisioned in this paper.

\begin{figure*}[!t]
\centering
\includegraphics[width=7in]{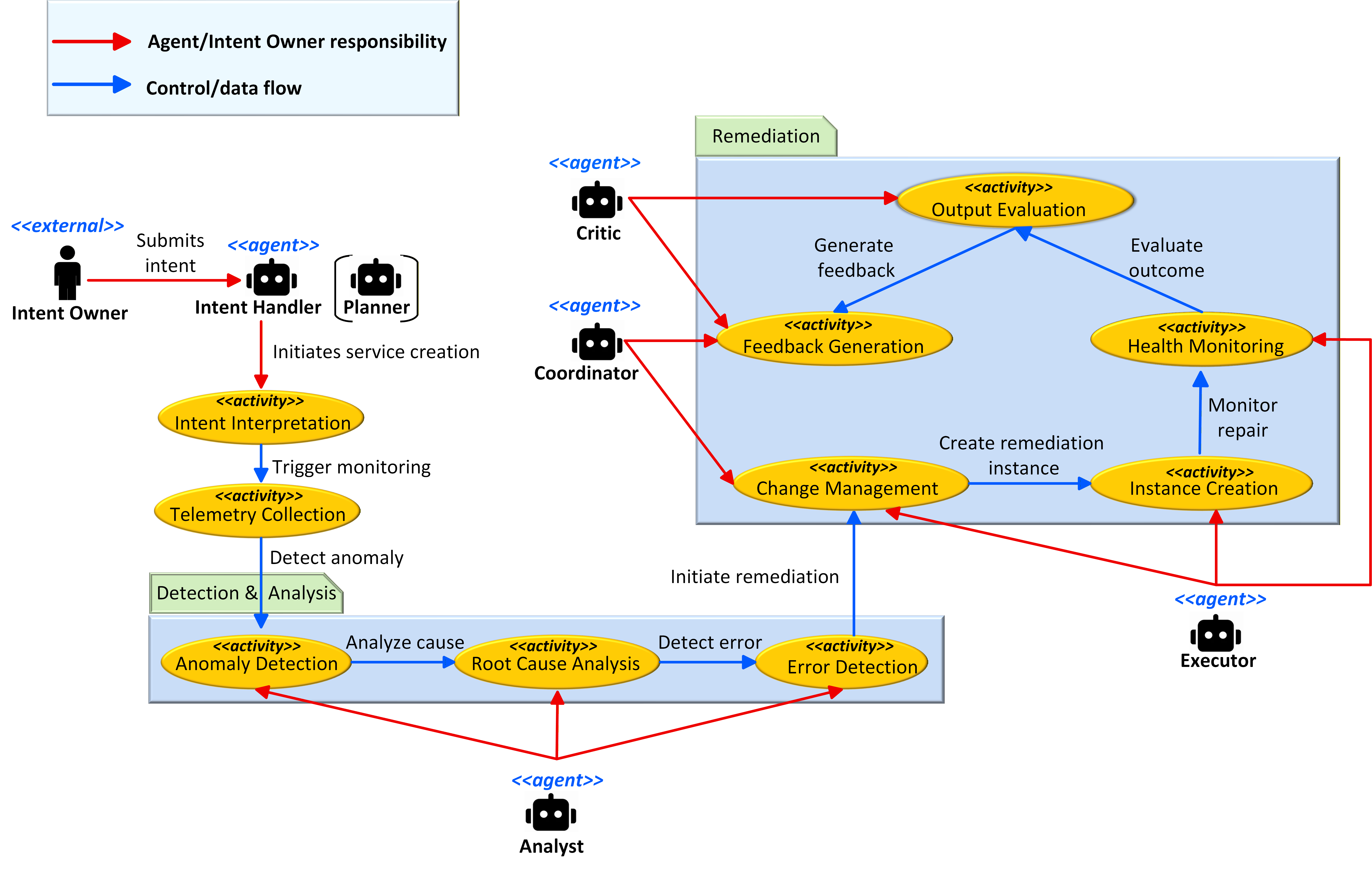}
\caption{\textcolor{black}{Fault management workflow showing how \emph{Planner}, \emph{Analyst}, \emph{Executor}, \emph{Coordinator}, and \emph{Critic} agents collaborate through intent interpretation, analysis, remediation, and feedback phases to autonomously resolve network faults. Red arrows indicate agent/intent-owner responsibilities and coordination relationships, while blue arrows represent the workflow progression and information exchange between activities.}}
\label{figures:hvs08-fault-management-usecase-diagram}
\end{figure*}

True autonomy means that each scenario depends on agents that interpret intent, coordinate resources, and adapt to changing conditions. Today, these agents are guided components with human-in-the-loop oversight. Future deployments could use recursive components for greater autonomy.
{\color{black}
\subsection{Use Case: Fault Management}

Returning to the electricity over-consumption example (Section~\ref{think}), we examine how autogenic network management uses agents to handle MAC scheduler drift, which is causing the excessive energy use (Fig.~\ref{figures:hvs08-fault-management-usecase-diagram}). The operator observes elevated electricity usage and expresses this concern through an intent such as \emph{``Reduce excessive RAN energy consumption while preserving service quality"}, specifying the desired outcome without prescribing how to achieve it.}

The fault management workflow begins when the Intent Handler, which is also the Planning agent, receives this intent, translates the high-level sustainability concern into a specific operational objective (self-orients), and plans what to do. In response, the Management system dynamically creates specialized agents (self-architects) and synthesizes the agent structure required for this fault scenario (self-programs).

Fig.~\ref{figures:hvs08-fault-management-usecase-diagram} shows the ordered workflow of agent actions and their interactions:
\begin{itemize}
\item The \emph{Planner (Intent Handler)} interprets fault intents and initiates the remediation workflow.
\item The \emph{Analyst} diagnoses service degradation through anomaly detection, root-cause analysis, and error identification.
\item The \emph{Executor} applies remediation actions, manages configuration changes, and monitors recovery progress.
\item The \emph{Coordinator} synchronizes change management and feedback activities across agents and network domains.
\item The \emph{Critic} validates remediation outcomes, confirms service restoration, and generates feedback for continuous improvement.
\end{itemize}

The workflow unfolds in phases. During intent interpretation and telemetry collection, the \emph{Planner} agent initiates continuous monitoring of network health. The \emph{Analyst} agent then performs root cause analysis and error detection to identify the fault. For change management and remediation, the \emph{Executor} and \emph{Coordinator} agents collaborate to implement tailored remediation actions. The \emph{Executor} agent tracks repairs in real time to confirm services meet baseline requirements. Finally, the \emph{Critic} and \emph{Coordinator} agents evaluate outcomes and generate feedback that informs future fault management strategies (self-reflects), enabling the system to learn and adapt over time.

As this use case is only one of 11 high-value scenarios (as described in Table \ref{table:high-value-scenarios-for-autonomous-networks} and Fig. \ref{figures:high-value-scenarios-usecase-diagram}) and represents an idealized resolution, the range of agents implies that network management must support a diverse set of agents, each with specialized roles and responsibilities.
It is unrealistic to expect a single vendor to supply all software for agents to carry out complex tasks and collaborate. Interoperability between vendors to enable autogenic capabilities will require standards to define common interfaces or other operational principles between agents and their environment.

\subsection{Transitioning from Guided to Recursive Components}

The fault management scenario (Fig. \ref{figures:hvs08-fault-management-usecase-diagram}) demonstrates how guided components operate today. The \emph{Planner} interprets intent and generates remediation plans using self-programming capabilities. However, humans remain in the loop. Operators validate the generated procedural logic before deployment, review configuration changes, and approve the remediation strategy. This human oversight provides safety boundaries but creates a bottleneck when handling multiple simultaneous faults across network domains.

The transition to recursive components can be achieved through solution banking. Validated solutions generated by guided components are stored in a trusted repository, allowing recurring fault scenarios to be handled autonomously based on proven solution patterns. Novel faults still require human oversight, but as the repository grows and validation mechanisms mature, the system accumulates reliable recovery strategies and progressively shifts from guided to recursive operation.

\section{Challenges and Research Roadmap}\label{section:challenges-and-roadmap}

Recursive components are an active research frontier. The transition from guided to recursive components requires advances in the four autogenic capabilities.

\subsection{Architectural Challenges and Research Directions}

Realizing autogenic network management requires fundamental changes to network software architecture and advances in AI capabilities. At the architectural level, the primary challenge is redesigning network software to support agent-based operation, where autonomous components reason about system state, generate actions, and validate their outcomes at runtime.

A key architectural enabler is the use of \emph{digital twins} as safe environments for evaluating autonomously generated control strategies. While digital twins reduce operational risk, their cost motivates the concept of a \emph{digital twin factory}. This is a special type of agent within the management subsystem (Fig.~\ref{figures:autogenic-system-reference-architecture}) that uses LAMs to interpret semantic models and automatically synthesize models and middleware required to acquire network data. The digital twin factory enables a flexible approach to define models of network resources and services, which can lead to search strategies to find high-fidelity digital twins for production networks even as configurations, workloads, and control logic evolve.

\textcolor{black}{A second architectural and security challenge} lies in \emph{interface design}. Autogenic management requires interfaces through which agents interact with network functions, yet poorly designed interfaces can amplify the risk of unsafe or unintended behavior. The research direction here is to define interfaces that are intentionally simple and constrained, making them difficult to misuse. Such interfaces rely on formal specifications that bound permissible operations, regulate resource access, and restrict valid state transitions. By enforcing these constraints, the verification burden shifts from analyzing arbitrary agent-generated logic to checking compliance with well-defined protocol specifications. \textcolor{black}{From a security perspective, these constraints reduce the potential impact of erroneous agent-generated actions by requiring authorization, validation, and rollback mechanisms before changes affect production network functions. This is particularly important for recursive components, which may generate or modify operational logic with reduced human supervision.}

These architectural considerations raise a key standardization challenge: defining interface abstractions that are expressive enough to support autonomous management while remaining sufficiently constrained to preserve safety and operational correctness. However, architecture alone is insufficient. Progress also depends on advances in AI capabilities, many of which are already being explored in adjacent research domains but not yet integrated into network management. The following subsection reviews state-of-the-art AI techniques that exhibit these autogenic capabilities, highlighting both their promise and limitations for supporting autogenic network management.

\subsection{State of the Art in AI Capabilities for Autogenic Systems}

Recent advances in AI research demonstrate partial realizations of the capabilities required for autogenic network management. We organize the state of the art around four core autogenic functions: self-programming, self-reflection, self-architecting, and self-orienting.

Self-programming has advanced through generative AI, particularly Large Language Models (LLMs) such as Claude \cite{anthropic2024claude}, which can synthesize code from natural-language intent to automate tasks with limited human input. FunSearch \cite{velickovic2024amplifying} further combines LLMs with evolutionary algorithms to discover novel solutions, moving toward open-ended creation. However, reliability and domain adaptation remain open challenges.

Self-reflection supports safer autonomy by letting systems evaluate and improve their outputs. As an example, the CRITIC framework \cite{gou2023critic} combines LLMs with external verification tools to enhance reasoning and reduce errors. Reflexion \cite{shinn2023reflexion} shows agents using episodic memory and self-generated critiques to improve decision-making without retraining. Despite these advances, LLMs alone do not provide sufficient self-reflection; external feedback and robust evaluation still remain essential.

Self-architecting enables systems to modify internal structure and logic. Neural Architecture Search (NAS) \cite{liu2019darts} enabled automated model design, while EvoMAC \cite{hu2024selfevolving} extends architectural evolution to multi-agent systems through dynamic reorganization. Although these approaches support runtime adaptation, architectural changes can disrupt functionality and therefore require rigorous safety frameworks.

Self-orienting shifts systems from executing predefined goals to generating objectives. The Darwin Gödel Machine \cite{zhang2025darwin} demonstrates open-ended evolution through emergent goal generation, while the Self-Taught Optimizer (STOP) \cite{zeilkman2024selftaught} uses language models as meta-optimizers to discover improvement strategies. However, autonomous goal-setting raises alignment, stability, and safety concerns, requiring mechanisms that keep objectives consistent with human values.

\section{Conclusions and Future Work}\label{Conclusion}

AI-native 6G networks require management systems capable of generating and evolving automation at runtime. We presented autogenic network management as the foundation for LAM-enabled self-evolving networks. Our proposed seven-subsystem architecture shows how programmed, learning, guided, and recursive components can be organized to realize system-level capabilities: self-orienting, self-programming, self-reflection, and self-architecting.

The architecture shows a realistic, standards-compatible deployment path. Operators can begin with guided components that use human-in-the-loop validation to create automation software safely. Solution banking enables a gradual transition toward recursive components as validated solutions accumulate and confidence in specific patterns grows. This staged approach manages risk while delivering the self-programming benefits needed for L4 autonomy.

Realizing autogenic systems requires progress on multiple fronts. 
Future work should strengthen the role of digital twins as safe testing environments where autonomously generated actions can be evaluated before deployment in production networks. Building on this, digital twin factories should be explored as management-plane agents that synthesize, update, and select appropriate twins for evolving network conditions. A second direction is the design of constrained, standards-compatible agent interfaces that define permissible operations, resource-access boundaries, valid state transitions, and rollback mechanisms. Finally, recursive components require stronger self-reflection and self-architecting capabilities, supported by external validators, trusted solution repositories, and large-scale validation across RAN, Core Network, and cross-domain service scenarios.
These challenges represent active research areas where progress is visible, but significant work remains.
\balance

\bibliographystyle{IEEEtran}

\section*{Biographies}
\vspace{-3em}
\begin{IEEEbiographynophoto}{Petar Djukic}
(petar.djukic@nokia-bell-labs.com) received the B.A.Sc., M.A.Sc., and Ph.D. degrees in computer engineering from the University of Toronto, Toronto, ON, Canada, in 1999, 2003, and 2007, respectively. He is currently with Bell Labs Research, Ottawa, ON, Canada. He holds over 65 U.S. patents and has authored 44 peer-reviewed publications. His research interests include multi-agent AI systems, autonomous networks, and reliable machine learning systems.
\end{IEEEbiographynophoto}

\begin{IEEEbiographynophoto}{Sudipta Acharya}
(sacharya2@uottawa.ca) received the Ph.D. degree in computer science and engineering from the Indian Institute of Technology (IIT) Patna, India, in 2019. She is currently a Postdoctoral Fellow at the University of Ottawa and a MITACS Accelerate Researcher at Nokia Bell Labs, Canada. Her research interests include deep learning and optimization, agentic AI, network digital twins, and AI-native 6G network intelligence.
\end{IEEEbiographynophoto}
\begin{IEEEbiographynophoto}{Takai Eddine Kennouche}
(takai.kennouche@nokia.com) received the Ph.D. degree in electrical and electronics engineering from the Università degli Studi di Pavia, Pavia, Italy. He is currently a Lead AI and Data Architect with Nokia Standards, Research and Innovation, France. Previously, he was a Lead AI Architect with VIAVI Solutions and a Postdoctoral Researcher with Inria, Grenoble. His research interests include AI-native network architectures, data governance for 6G, MLOps, and telecom automation.
\end{IEEEbiographynophoto}

\begin{IEEEbiographynophoto}{Burak Kantarci}
[S'05, M'09, SM'12] (burak.kantarci@uottawa.ca) is a Full Professor and University Research Chair in AI-Enabled Secure Networking at the University of Ottawa, where he also directs the Smart Connected Vehicles Innovation Centre and the NEXTCON Lab. He serves as an editor for several IEEE journals and is a Distinguished Lecturer of both the IEEE Communications Society and IEEE Systems Council.
\end{IEEEbiographynophoto}
\end{document}